\documentclass[aps,prb,twocolumn, superscriptaddress,showpacs,floatfix]{revtex4-1}
\usepackage{float}
\usepackage{graphicx}
\usepackage{amsmath}
\usepackage{amssymb}
\usepackage{url}
\usepackage{color}
\usepackage{natbib}

\usepackage{bm}

\graphicspath{{.}{../}{./eps/}}

\renewcommand{\d}[2]{\frac{d #1}{d #2}} 

\newcommand{\ket}[1]{\left| #1 \right>} 
\newcommand{\bra}[1]{\left< #1 \right|} 
\let\baraccent=\= 
\renewcommand{\=}[1]{\stackrel{#1}{=}} 

\newcommand{\eq}[1]{(\ref{#1})}

\newcommand{\medio}[1]{\left\langle #1 \right\rangle}

\def\XXint#1#2#3{{\setbox0=\hbox{$#1{#2#3}{\int}$}
     \vcenter{\hbox{$#2#3$}}\kern-.5\wd0}}

\renewcommand{\exp}[1]{\text{e}^{#1}}

\begin{document}

\title{ Kubo-Bastin approach for  the spin Hall conductivity of decorated graphene }

\author{Jose H. Garcia}
\affiliation{ICTP South American Institute for Fundamental Research,
Instituto de F \'{\i}sica Te\' orica, UNESP-Universidade Estadual Paulista, 01140-070, Sao Paulo, SP, Brasil}
\author{Tatiana G. Rappoport}
\affiliation{Instituto de F\'\i sica, Universidade Federal do Rio de Janeiro, Caixa Postal 68528, 21941-972 Rio de Janeiro RJ, Brazil}

\date{\today}

\begin{abstract}

Theoretical predictions and recent experimental results suggest one can engineer spin Hall effect in graphene by enhancing the spin-orbit coupling in the vicinity of an impurity. We use a Chebyshev expansion of the Kubo-Bastin formula to compute the spin conductivity tensor for a tight-binding model of graphene with randomly distributed impurities absorbed on top of carbon atoms. We model the impurity-induced spin-orbit coupling with a graphene-only Hamiltonian that takes into account three different contributions~\cite{Gmitra2013}  and show how the spin Hall and longitudinal conductivities depend on the strength of each spin-orbit coupling and the concentration of impurities. Additionally, we calculate the real-space projection of the density of states in the vicinity of the Dirac point for single and multiple impurities and correlate these results with the conductivity calculations.
\end{abstract}


\maketitle

\section{Introduction}
Several non-magnetic materials with strong spin-orbit coupling present electrically induced transverse spin-currents in the absence of an external magnetic field. This  is commonly referred as  spin Hall effect (SHE) because of its similarity with the regular charge Hall effect \cite{Dyakonov2008, Kato2004, Sih2005}. This class of materials may play an important role in spintronics, creating  electrically controlled spin-polarized currents in non-magnetic systems, which can be used for spin injection\cite{Jungwirth2012}.

The spin-orbit coupling (SOC) in graphene is extremely weak, making the detection of the spin Hall effect impossible. However, its high mobility and long spin relaxation times make it a good candidate as spin conductors in spintronics~\cite{Han2014}.  It has been proposed that the functionalization of graphene with adatoms can produce a quantum spin hall state (QSHE) ~\cite{Kane2005} if the induced spin-orbit coupling (SOC) has the same symmetry of the intrinsic spin-orbit in graphene~\cite{weeks2011}. The observation of this state is experimentally challenging: adatoms can form clusters, can introduce additional SOCs and can break symmetries, destroying the topological state~\cite{Jiang2012, Cresti2014}. However, even without a quantized spin hall conductivity, it is still possible to use both intrinsic and Rashba  SOC to generate spin Hall effect in graphene with large spin Hall angles~\cite{Ferreira2014}. Recent experiments show that different adsorbed adatoms and metallic clusters induce a giant spin-orbit enhancement in graphene, \cite{Balakrishnan2013, Balakrishnan2014, Avsar2015, Calleja2015}, opening the possibility to use induced spin-orbit couplings to create active graphene-based spintronic devices.

Balakrishnan {\it et al.} were the first to observe this effect experimentally. They reported that weakly hydrogenated \cite{Balakrishnan2013} and fluorinated \cite{Avsar2015} graphene presented an enhancement of the spin-orbit coupling of more than one order of magnitude with respect to pristine graphene, giving rise to spin Hall effect.  It is believed that the increase in the spin-orbit coupling in light adatoms is attributed to the transformation of the $sp^2$ bonds into $sp^3$, breaking the mirror symmetry in graphene and inducing a local spin-orbit coupling \cite{AHCN2009}. Although the detailed physical mechanism of the enhancement of the  spin-orbit coupling is still unclear, a first advance toward its understanding was the minimal effective tight-binding hamiltonian proposed in references \onlinecite{Gmitra2013,Irmer2015}, where they separate, with the help of group theory, the spin-orbit splitting obtained in DFT calculations into different contributions:  the well known intrinsic spin-orbit (ISO) and Rashba  
couplings (RSO) and a novel contribution referred as pseudospin inversion asymmetry (PIA). Other effective tight-binding hamiltonians that take into account additional terms and/or deal with the adsorption of atoms in other positions had also been proposed \cite{weeks2011,Pachoud2014}.

\hspace{5mm}
Recently, we developed a a real space implementation of the Kubo formalism to calculate the electronic conductivity tensor of large systems at finite temperatures~\cite{Garcia2015} where we expand the Kubo-Bastin formula~\cite{Bastin1971b}  in terms of Chebyshev polynomials~\cite{Silver1996,Weisse2006}. 
In this article, we extend the Kubo-Bastin formula to calculate the spin conductivity tensor and apply this modified Kubo formula to calculate the spin Hall conductivity of graphene decorated with adatoms. We consider a minimal tight-binding Hamiltonian that models the spin-orbit interaction of adatoms sitting at the $T$-site~\cite{Konschuh2010,Gmitra2013,Irmer2015} and  we calculate the spin and charge quantum transport of the system. We perform a systematic analysis of  the contribution of  each term of the effective Hamiltonian to the spin Hall effect for different concentrations and couplings. We also examine the real-space local density of states to a gain a further insight on these mechanisms.

The article is organized as follows: in Section II we present the effective tight-binding Hamiltonian for graphene with spin-orbit adatoms sitting at the $T$-site. 
In section III we discuss our numerical approach to calculate the charge conductivity tensor with the Chebyshev expansion of the Kubo-Bastin formula. We also present its extension to calculate the spin conductivity tensor.  In section IV we present a systematic numerical analysis of the charge and spin conductivity as a function of concentration and spin-orbit coupling and the local density of states for the different types of spin-orbit couplings. Finally, in sec. V we present our conclusions.

\section{Spin-Orbit Coupling in Graphene with adatoms}

A minimal tight-binding model for adatoms adsorbed at the $T$ site has been proposed recently \cite{Gmitra2013,Irmer2015} where they consider a Hamiltonian, based on the non-interacting one impurity Anderson Hamiltonian, of the form:
\begin{equation}
H=H_g+\epsilon_0\sum_\sigma{f^\dagger_\sigma f_\sigma}+ V\sum_\sigma{(a^\dagger_{I\sigma}f_\sigma + h. c.)} +H_{SO}.\label{GMITRA}
\end{equation}
 $a^\dagger_I$ and $a_I$ are the creation and annihilation operators of the carbon atom where the adatom is adsorbed, which in this case is defined as belonging to the $A$ sub-lattice.   $\epsilon_0$ is the energy of the localized state of the adatom and $ f$ is the annihilation operator for the localized electrons. The third term describes the hybridization between the impurity and graphene and $H_{SO}$ is a graphene only term that takes into account the local modification of the spin-orbit coupling in the presence of the adatom;  its form is determined through group theory. The adatom degrees of freedom in Hamiltonian Eq.\eq{GMITRA} can be integrated out and in the vicinity of to the Dirac point, the impurity acts as a local potential of strength  $\epsilon_{I}=-V^2/\epsilon_0$ \cite{Uchoa2009a}.

The adsorption of adatoms at the $T$-site locally brakes the sub-lattice and out-of-plane mirror symmetry, allowing the existence of different spin-orbit terms. In reference \onlinecite{Gmitra2013},  they perform DFT calculations of weakly hydrogenated graphene and fit their data with the model of Eq.\eq{GMITRA}. From their analysis, only three different spin-orbit contributions are necessary to fit the data:
\begin{equation}
H_{SO}=H_{I}+H_{R}+H_{PIA}\label{SpinOrbitHamiltonian}
\end{equation}
where $H_{ISO}$ is the local intrinsic spin-orbit coupling given by
\begin{equation}
H_{I}=\frac{i\lambda_{I}}{3\sqrt{3}}\sum_{i,j \in C_{I}} b_{i,\sigma}^\dagger \nu_{i,j}(\hat{s}_z)_{\sigma,\sigma}b_{j,\sigma} +h. c.
\end{equation}
where $C_{I}$ represents the set of nearest-neighbors of the carbon atom at the $\bm{R}_I$ position  where the impurity was adsorbed, $\sigma_i$, $i=x,y,z$ are the Pauli matrices acting on the spin $\sigma$ and $\nu_{i,j}=\pm1$ is dictated by the orientation of the hopping processes (either clockwise or counterclockwise).   $H_{R}$ represents a local Rashba coupling that induces spin-flip hopping between the carbon atom at $\bm{R}_I$  and its nearest neighbors :
\begin{equation}
H_{R}=\frac{2i\lambda_{R}}{3}\sum_{\medio{I,j},\sigma}a_I\sigma^\dagger (\hat{\sigma}\times d_{I,j})_{z,\sigma,\bar{\sigma}}b_{j,\bar{\sigma}} +h. c.
\end{equation} where $d_{I,j}$ are the distances  from $\bm{R}_I$  to the three nearest  carbon atoms. 

\begin{figure}
\centering
\includegraphics[width=1.0\linewidth,clip]{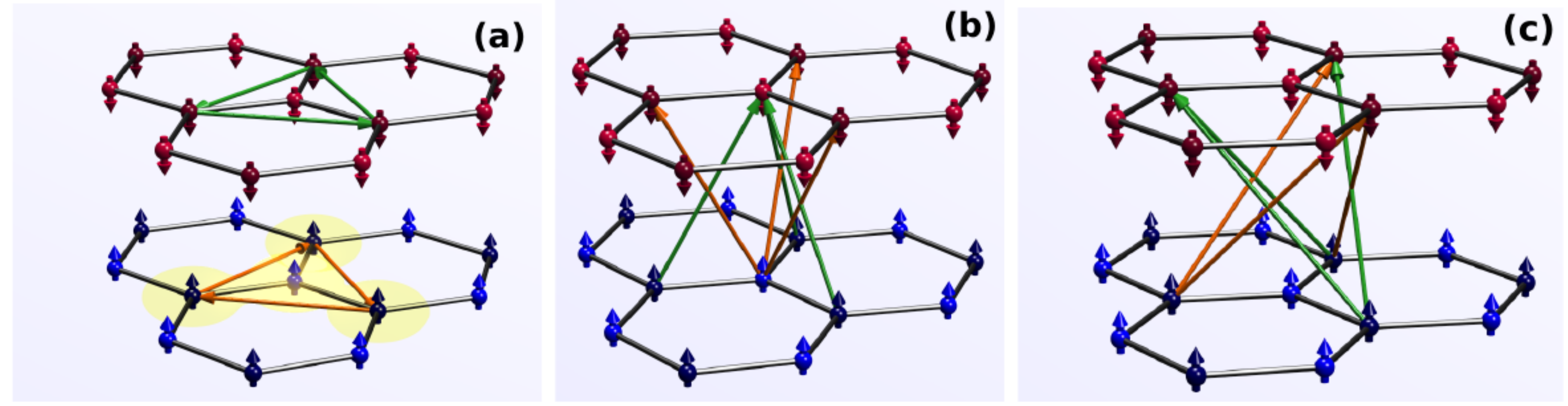}
\caption{ Representation of the local spin-orbit couplings generated by the adsorption of adatoms in the $T$-site. (a) Local intrinsic spin-orbit coupling, (b) Local Rashba spin-orbit coupling and (c) Local pseudo-spin inversion asymmetry. }
\label{DifferentSpinOrbit}
\end{figure}

There is a third term that is related to the $C_{3v}$ symmetry that emerges when an adatom is adsorbed in the top position. It describes spin-flip hoppings between next-nearest neighbors:
\begin{equation}
H_{PIA}=\frac{2i\lambda_{PIA}}{3}\sum_{\medio{\medio{i,j}},\sigma}b_{i,\sigma}^\dagger(\hat{s}\times D_{j,i})_{z,\sigma,\bar{\sigma}}B_{j,\bar{\sigma}} .
\end{equation}
$H_{PIA}$ is referred as pseudospin inversion asymmetry (PIA), which originates from  the broken sub-lattice symmetry that makes the sites in $A$ and $B$ sublattices inequivalent in the vicinity of the adatom. This term basically connects next nearest neighbors site with opposite spins. A schematic visualization of the three terms is given in Fig. \ref{DifferentSpinOrbit}
Several $T$-site adatoms had been examined and these contributions seem enough to fit the DFT calculations. However, it is important to notice that the minimum model used to obtain the three different contributions to the spin-orbit coupling does not take into account inter-valley scattering. If inter-valley is also considered, the symmetries of  Hamiltonian allow for more contributions to the effective spin-orbit coupling~\cite{Pachoud2014,Mahmoud2015}.

\section{Chebyshev expansion of the Kubo-Bastin formula}

The standard Kubo's formula for the conductivity tensor is a current-current correlation function \cite{Kubo1957,Mahan2000}
\begin{equation}
\sigma_{\alpha,\beta}(\mu,T)=\frac{1}{\Omega}\int_0^\infty dt\int_0^{\infty}d\lambda \medio{f(\mu,T,H) j_\alpha j_\beta(t+i\hbar\lambda)}\label{OriginalKuboFormula}
\end{equation}
where $j_\alpha$ is the electronic current operator, $f(\mu,T,H)\equiv 1/(1+\exp{-(\mu-H)/k_bT})$ the Fermi-Dirac distribution for a given chemical potential $\mu$ and temperature $T$ and $\Omega$ is the volume of the sample. This formula was deduced under very general conditions, requiring only that the electric field is weak enough for the linear response theory to be valid. In many cases, the electron-electron interaction is weak and one can rewrite Eq.\eq{OriginalKuboFormula} in the non-interacting electron approximation as the Kubo-Bastin formula \cite{Bastin1971b}:
\begin{align}
&\sigma_{{\alpha\beta}}(\mu, T)=\frac{ie^2\hbar}{\Omega}
\int_{E^-}^{E^+}d\varepsilon f(\varepsilon) \times  \label{KuboBastinFormula}\\
&\text{Tr}\left[v_\alpha\delta(\varepsilon-H)v_\beta \d{G^{+}({\varepsilon})}{\varepsilon}-v_\alpha \d{G^{-}(\varepsilon)}{\varepsilon}v_\beta \delta(\varepsilon-H)\right ], \nonumber 
\end{align}
where $H$ and $v_\alpha$ are the non-interacting electron Hamiltonian and velocities operators.  In real space, the velocity operator $\bm{v}$, can be expressed in terms of the eigenvalues of the position operator $\ket{\bm{R}_i},~i=1,\dots D$ and the Hamitonian's matrix elements by using the Heisenberg relation
\begin{equation}
\bm{v}=\frac{1}{i\hbar}[\bm{R},H]=\frac{1}{i\hbar}\sum_{i,j=1}^D (\bm{R}_i-\bm{R}_j)H_{i,j}\ket{\bm{R}_i}\bra{\bm{R}_j}.
\end{equation}
This particular form is very useful for tight-binding Hamiltonians, where in general the value of matrix elements connecting distant sites is zero.

Our method, presented in  Ref~\onlinecite{Garcia2015} and applied to quantum transport calculations of different solid state systems~\cite{Liu2015, Yuan2015} consists in expanding the Green's functions in the integrand of eq.(\ref{KuboBastinFormula}) in terms of Chebyshev polynomials using the kernel polynomial method~\cite{Silver1996,Weisse2006}.
The first step to approximate the Green's functions in terms of the Chebyshev polynomials is to rescale the energy spectra into the domain of the polynomials $ [-1,1] $. For numerical calculations, it is recommended to avoid an approximation near the edges of the interval. Therefore, we choose to rescale the Hamiltonian within the interval $[-\alpha,\alpha]$, with $\alpha\in(0,1)$ being a given positive parameter. This can be done through the following linear transformation
\begin{align}
\tilde{H}=\frac{2\alpha}{\Delta E}\left(H -\frac{E^{+}+E^{-}}{2}\right), \nonumber \\
\tilde{\varepsilon}=\frac{2\alpha}{\Delta E}\left(\varepsilon -\frac{E^{+}+E^{-}}{2}\right),
\end{align}
where $E^ {-} $ and $E^ {+} $ are the minimum and maximum eigenvalue of the spectrum and $\Delta E\equiv E^ {+} -E^ {-} $. In this article, we choose $\alpha=0.9$. To estimate the bounds, we apply the power method~\cite{powerbook}, which is normally used to locate dominant eigenvalues in linear algebra. The rescaled Hamiltonian and energy are  represented by $\tilde{H}$ and $\eta$ respectively and we can expand the rescaled delta and Green´s functions by considering their spectral representations $\delta(\eta-\tilde{H})=\sum_{k}\ket{E_k}\bra{E_k}\delta(\eta-\tilde{E}_k)$ and $G^{\pm}(\eta,\tilde{H})=\sum_{k}\ket{E_k}\bra{E_k}G^{\pm}(\varepsilon,\tilde{E}_k)$ and by expanding their eigenvalues in terms of the Chebyshev polynomials:
\begin{align}
&\delta(\tilde{\varepsilon}-\tilde{H})=\frac{2}{\pi \sqrt{1-\tilde{\varepsilon}^2}}\sum_{m=0}^M g_m\frac{T_m(\tilde{\varepsilon})}{\delta_{m,0}+1}T_m(\tilde{H}), \\
& G^\pm(\tilde{\varepsilon},\tilde{H})=\mp \frac{2i}{\sqrt{1-\tilde{\varepsilon}^2}}\sum_{m=0}^M g_m\frac{\exp{\pm i m \text{arccos}(\tilde{\varepsilon})}}{\delta_{m,0}+1} T_m(\tilde{H}).
\end{align}
where {$T_m(x)=\cos[m\arccos(x)]$} is the Chebyshev polynomial of the first kind and order
$m$, which is defined according to the recurrence relation
{$T_{m}(x)=2xT_{m-1}(x)-T_{m-2}(x)$}. The expansion has a finite number of terms ($M$) and  the
truncation gives rise to Gibbs oscillations that can be smoothed with the use of a kernel,
given by $g_m$~\cite{Silver1996,Weisse2006}.

Replacing the expansions above in eq.\eq{KuboBastinFormula} with $\Delta E=E^{+}-E^-$,  we obtain
\begin{align}
\sigma_{{\alpha\beta}}(\mu, T)&= \frac{4e^2\hbar}{\pi \Omega}\frac{4}{\Delta E^2}
\int_{-1}^{1}d\tilde{\varepsilon}
\frac{f(\tilde{\varepsilon})}{(1-\tilde{\varepsilon}^2)^2}\sum_{m,n}\Gamma_{{nm}}
(\tilde { \varepsilon } )\mu^{\alpha\beta}_{{nm}}\label{condfinal}
\end{align}
where $\mu^{\alpha\beta}_{{mn}}\equiv \frac{g_m
g_n}{(1+\delta_{{n0}})(1+\delta_{{m0}})}\text{Tr}\left[v_\alpha T_m(\tilde{H})v_\beta
T_n(\tilde{H})\right]$ does not depend on the energy. Since $\mu_{{mn}}$ involves products of
polynomial expansions of the Hamiltonian, its calculation is responsible for most of the 
computational cost.

On the other hand, {$\Gamma_{mn}(\tilde{\varepsilon})$} is a scalar that is energy dependent
but independent of the Hamiltonian
\begin{equation}
\begin{split}
{\Gamma_{mn}(\tilde{\varepsilon})\equiv[(\tilde{\varepsilon}-i
n\sqrt{1-\tilde{\varepsilon}^2})\exp{in\text{arccos}(\tilde{\varepsilon})}T_m(\tilde{\varepsilon})}
\\
+(\tilde{\varepsilon}+im\sqrt{1-\tilde{\varepsilon}^2})\exp{-im\text{arccos}(\tilde{\varepsilon})}T_n(\tilde{\varepsilon})].
\end{split}
\end{equation}
 $\mu_{m,n}$ are the expansion moments of the polynomial expansion and 
 $\Gamma_{m,n}(\tilde{\varepsilon})$ are the expansion functions. As shown in \eq{condfinal}, once the
  coefficients $\mu_{{mn}}$ are determined, we can obtain
the conductivities for all temperatures and chemical potentials without repeating the most
time-consuming part of the calculation~\cite{Weisse2004}. Moreover, the recursive relations between
Chebyshev polynomials lead to a recursive multiplication of sparse Hamiltonian matrices that can be
performed in a very efficient way in GPUs~\cite{Covaci2010,Harju2013}.  Instead of the full
calculation of traces, we use self-averaging properties, normally used in Monte-Carlo calculations,
to replace the trace in the calculation of  $\mu_{{mn}}$.  With this method, known as random phases vector approximation \cite{Weisse2006, Iitaka2004,Silver1996} , we construct a set of $R\ll N$ complex vectors
\begin{align}
\ket{r}\equiv(\xi^r_1 ,\dots,\xi^r_N),\quad r=1,\dots,R,
\end{align}
with dimension  equal to $N$ and whose elements $\xi^r_i$ are drawn from a probabilistic distribution with the following characteristics: 
\begin{align}
\medio{\medio{\xi^r_i}}=0,\quad
\medio{\medio{{\xi^r_i}^*\xi^{r'}_j}}=\delta_{r,r'}\delta_{i,j}\label{conditions},
\end{align}
where $\medio{\medio{\dots}}$ is the statistical average.  The trace can be calculated as the average expected value of this random vector.The conductivities are averaged over several disorder {realizations}, $S$, with $R$ random vectors for each of them. Because of the self-averaging properties of large systems, the product $SR$ is the main defining factor of the accuracy of the trace operation.

\subsection{The Kubo-Bastin formula for Spin-Conductivity}

The Kubo-Bastin formula presented in Ref. \onlinecite{Garcia2015}, was derived for  spinless non-interacting electrons. To study the spin Hall effect numerically, it is necessary to extend this formula to calculate the spin Hall conductivity $\sigma_{sH}$. This was done in details in Ref. \cite{Yang2006} by replacing the first velocity operator in Eq.\eq{KuboBastinFormula} by the spin-current operator $J_x^\gamma$ defined as:
\begin{equation}
J_x^\gamma=\frac{1}{2\Omega}\{s_\gamma,v_x\}.
\end{equation}
where $s_\gamma$ is the spin operator and $\gamma=x,y,z$. This substitution leads to the following Kubo-Bastin formula for Spin-Conductivity\cite{Yang2006}:
\begin{align}
&\sigma_{\alpha,\beta}^z=ie\hbar\int_{-\infty}^{\infty}d\varepsilon f(\varepsilon) \times \\
&\text{Tr}\medio{J_x^z\delta(\varepsilon-H)v_y\d{G^+(\varepsilon)}{\varepsilon}-J_x^z\d{G^-(\varepsilon)}{\varepsilon}v_y\delta(\varepsilon-H)} \nonumber.
\end{align}
Then, the Chebyshev approximation can be performed in an analogous way as the one described in Ref. \onlinecite{Garcia2015}, by modifying the moments
\begin{align}
\mu^{\alpha,\beta}_{m,n}\rightarrow \mu^{\alpha,\beta,\gamma}_{m,n}=\text{Tr}\left[T_m(H)J_\alpha^\gamma T_n(H)v_\beta \right].
\end{align}
A generalized Kubo formula for spin transport can also be obtained using non-abelian fields, which consider also external fields acting on the spin \cite{Jin2006}. However for the present purpose, the spin Kubo-Bastin formula is adequate.  

\section{Numerical calculations}

To gain an insight of the effect of the locally induced spin-orbit coupling in the spin and charge conductivities of graphene, it is necessary to analyze the role of the three different contributions of the spin-orbit coupling separately, that are referred here as ISO, RSO and PIA impurities. 
In this section, we perform a systematic analysis of the spin and charge conductivities of graphene in the presence of randomly distributed spin-orbit impurities as a function of the impurity concentration and the intensity of the coupling of each of the three contributions. For the numerical calculations, we consider a honeycomb lattice with $D=2\times N\times N$ with $N=200$ sites with periodic boundary conditions. For the statistical analysis we use $R=20$ random vectors and $S>40$ disorder realizations such that $SR>800$. All the conductivity calculations were performed using the same temperature $T=0.01t$. The systematic analysis as a function of the intensity was performed considering different values of the spin-orbit coupling at a fixed concentration of adatoms  $x=0.2$.
For simplicity, we do not consider any local potential in the carbon site where the adatom is located. 

\begin{figure}
\centering
\includegraphics[width=\linewidth,clip]{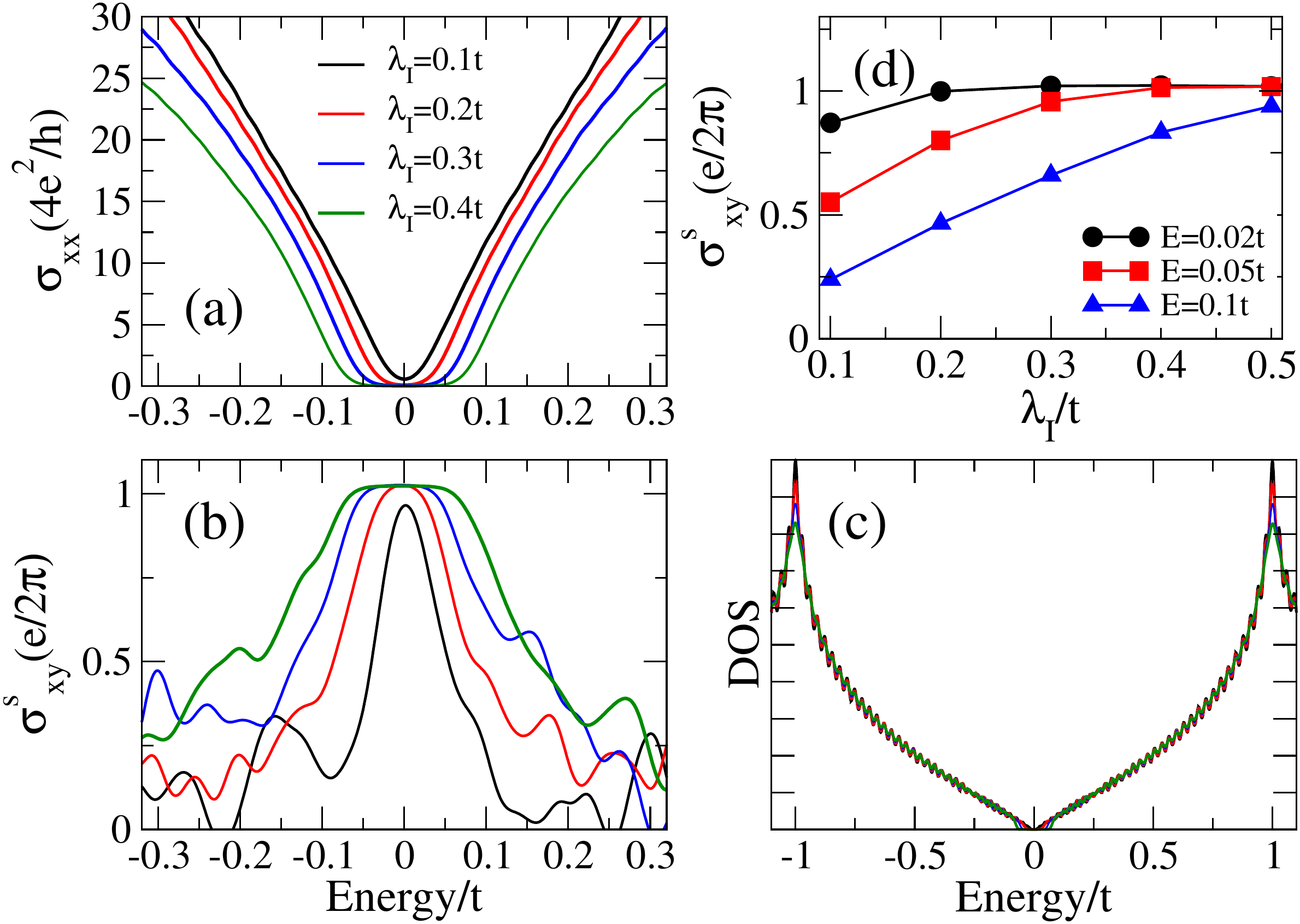}
\caption{ Graphene decorated with a random distribution of ISO-impurities ($\lambda_{R}=\lambda_{PIA}=0$) sitting at the $T$-site.  $D=2\times 200\times 200$ sites,  concentration of $x=0.2$ and  $M=1600$. (a) Longitudinal conductivity  (b) Spin Hall conductivity  as a function of the chemical potential for increasing values of the intensity of the ISO coupling. (c) Density of state  (d) Spin Hall conductivity as a function of the spin-orbit intensity for different values of the Fermi energy: (Black) 0.02t, (Red) 0.05t, (Blue) 0.1t. }
\label{Chap5-ISOINT}
\end{figure}

We begin  our analysis by considering the effect of $H_{I}$. In Fig.\ref{Chap5-ISOINT} we show the KPM numerical calculations with $M=1600$ moments for a honeycomb lattice decorated with a random distribution of pure intrinsic adatoms $(\lambda_{R}=\lambda_{PIA}=0)$. The density of states is shown in  Fig.\ref{Chap5-ISOINT}.c where the presence of a band gap is observed.  This gap translates into a region of zero longitudinal conductivity, shown in Fig.\ref{Chap5-ISOINT}.b.
\begin{figure}
\centering
\includegraphics[width=\linewidth,clip]{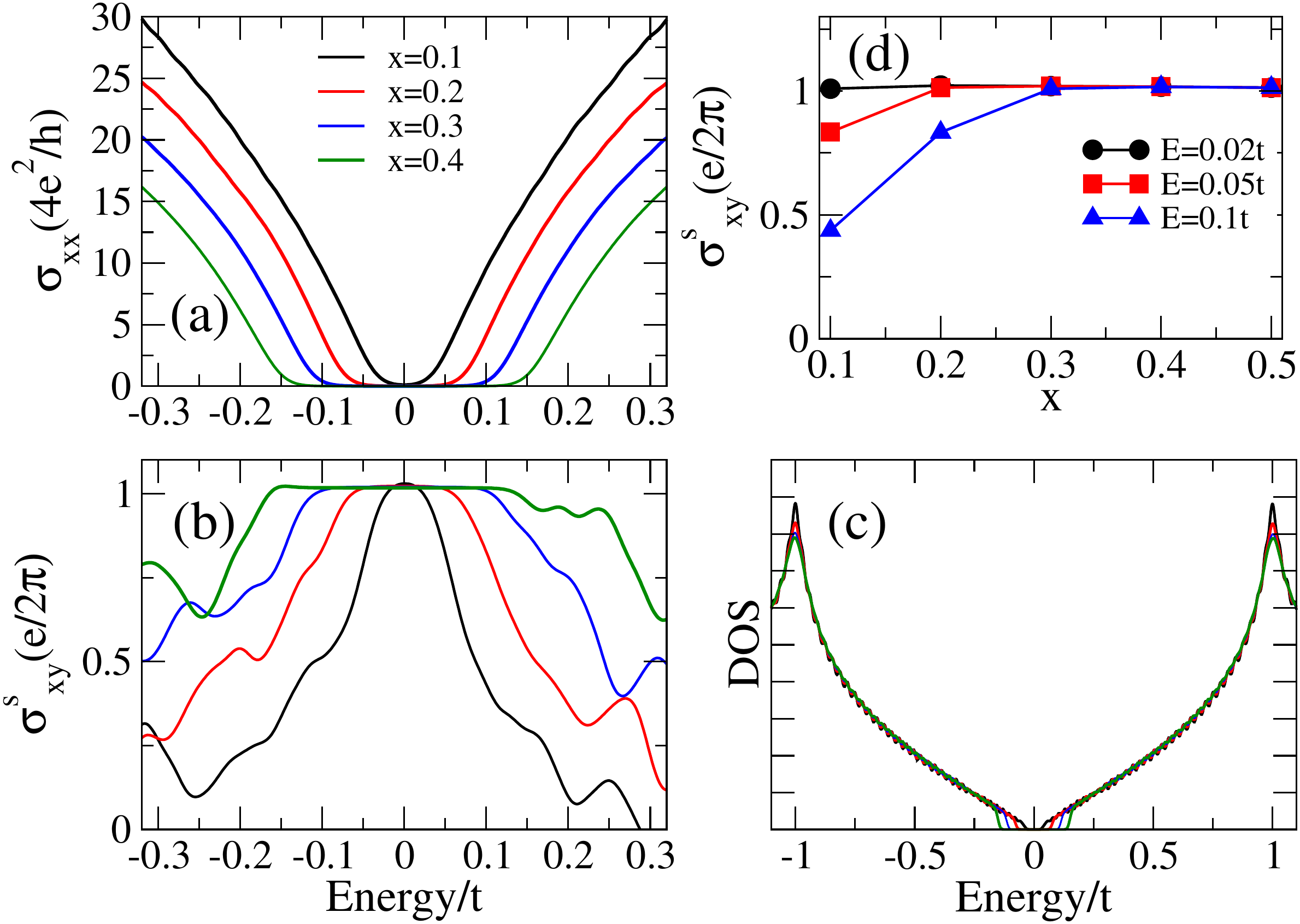}
\caption{ Graphene decorated with a random distribution of ISO-impurities ($\lambda_{R}=\lambda_{PIA}=0$) sitting at the $T$-site. $D=2\times 200\times 200$ sites,   $\lambda_I=0.4t$ and  $M=1600$. (a) Longitudinal conductivity  (b) Spin Hall conductivity  as a function of the chemical potential for increasing values of the concentration $x$. (c) Density of state  (d) Spin Hall conductivity as a function of the concentration for different values of the Fermi energy: (Black) 0.02t, (Red) 0.05t, (Blue) 0.1t. }
\label{Chap5-ISOCON}
\end{figure}
For this concentration, the value of the gap follows the following relation $$\Delta_{ISO}\propto x\lambda_{I},$$ which is consistent with Kane and Mele~\cite{Kane2005} model rescaled by the concentration and agrees with previous numerical calculations~\cite{Liu2015,Garcia2015}. For intensities below $\lambda_{I}=0.1t$, the gaps close due to the effect of the temperature. In Fig.\ref{Chap5-ISOINT}.c we can see that inside the gap, the system presents a quantized spin Hall conductivity as expected from the Kane and Mele model. However, we observe that outside the gapped region a robust spin Hall conductivity still persists, a fact that can be important in experiments where the intensity of the ISO coupling is usually small and the gap is closed. 

\begin{figure}
\centering
\includegraphics[width=\linewidth,clip]{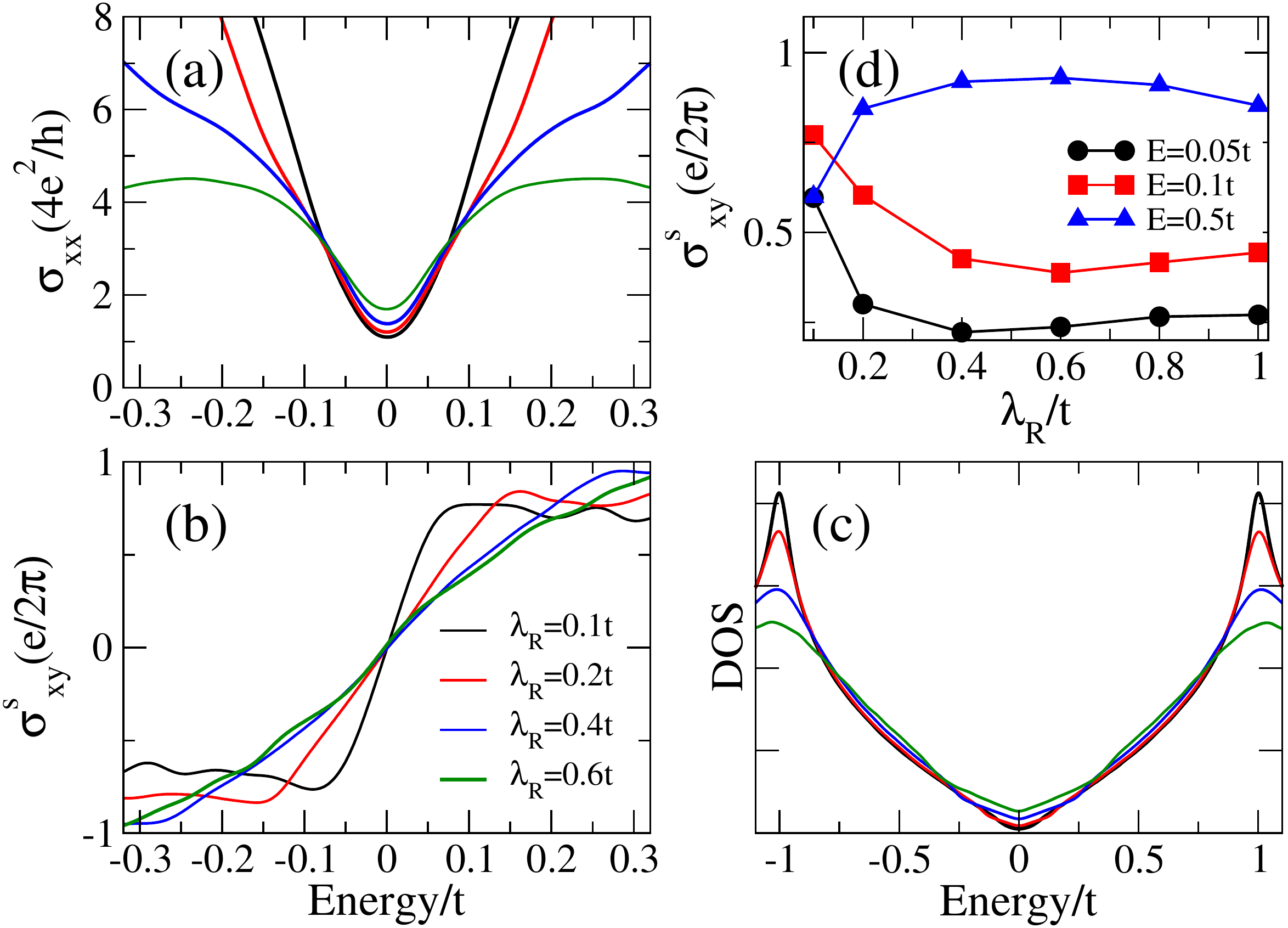}
\caption{ Graphene decorated with a random distribution of RSO-impurities ($\lambda_{I}=\lambda_{PIA}=0$) sitting at the $T$-site.  $D=2\times 200\times 200$ sites,  concentration of $x=0.2$ and  $M=800$. (a) Longitudinal conductivity  and (b) spin Hall conductivity  as a function of the chemical potential for increasing values of the intensity of the ISO coupling. (c) Density of state  (d) Spin Hall conductivity as a function of the spin-orbit intensity for different values of the Fermi energy: (Black) 0.05t, (Red) 0.1t, (Blue) 0.5t. }
\label{Chap5-RSOINT}
\end{figure}

In Fig.\ref{Chap5-ISOCON}, the same analysis is performed for a fixed coupling intensity and different concentrations. For these values of concentrations, the behavior is similar to the previous case. One can see a topological gap whose size scales as $\Delta_{I}\propto x\lambda_{I}$, with a robust spin Hall conductivity outside the gap. In experiments, the concentration is much smaller than the ones considered here, and this linear scaling of the gap with concentration might change below some critical concentration.
\begin{figure}
\centering
\includegraphics[width=\linewidth,clip]{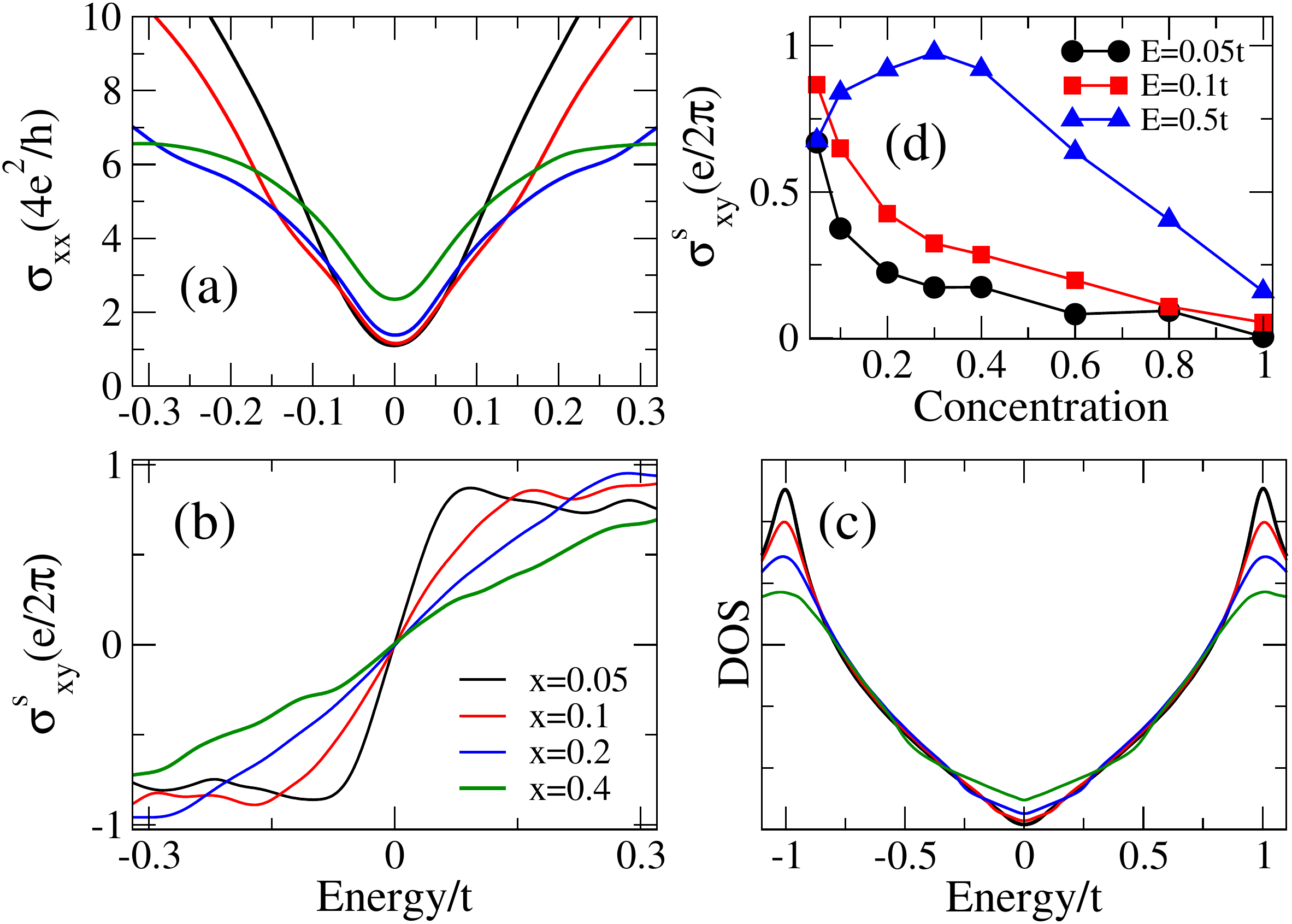}
\caption{ Graphene decorated with a random distribution of RSO-impurities ($\lambda_{I}=\lambda_{PIA}=0$) sitting at the $T$-site.  $D=2\times 200\times 200$ sites,  $\lambda_R=0.4t$ and  $M=800$. (a) Longitudinal conductivity  and (b) spin Hall conductivity  as a function of the chemical potential for increasing values of concentration $x$. (c) Density of state  (d) Spin Hall conductivity as a function of the sconcentration for different values of the chemical potential: (Black) 0.05t, (Red) 0.1t, (Blue) 0.5t. }

\label{Chap5-RSOCON}
\end{figure}
However, the robust spin Hall conductivity for both low concentration and low intensity, even in the absence of the topological gap, may play an important role in experimental results.

Our analysis for  a distribution of pure Rashba spin-orbit impurities ($\lambda_{I}=\lambda_{PIA}=0$) as a function of intensity of the coupling can be seen in Fig.\ref{Chap5-RSOINT}, where we keep the  concentration fixed ($x=0.2$). In the density of states shown in Fig.\ref{Chap5-RSOINT}.c, we notice the presence of new states at the neutrality point. These states are extended as can be seen in panel \ref{Chap5-RSOINT}.a with an slight increase in the minimum of the longitudinal conductivity for increasing values of the spin-orbit coupling. At the same time, the rashba coupling strongly suppress the longitudinal conductivity away from the charge neutrality point. The spin Hall conductivity changes signs with the energy and it is zero at the Dirac point, as expected (see Fig.\ref{Chap5-RSOINT}.c) . However in the vicinity of the $E=0$ there is a rapid increase of the spin Hall conductivity that saturates at $\approx\pm e/(2\pi)$, which is consistent with analytical calculations for the SHE in graphene with constant RSO ~\cite{Dyrdal2009}. The transition from negative to positive spin Hall conductivity as a function of the chemical potential gets more abrupt for weak $\lambda_{R}$. Surprisnly, this translates into an {\it increase} of the spin Hall effect in the vicinity of the neutrality point for {\it decreasing}  values of $\lambda_{R}$, as shown in Fig.\ref{Chap5-RSOINT}.d.

\begin{figure}
\centering
\includegraphics[width=\linewidth,clip]{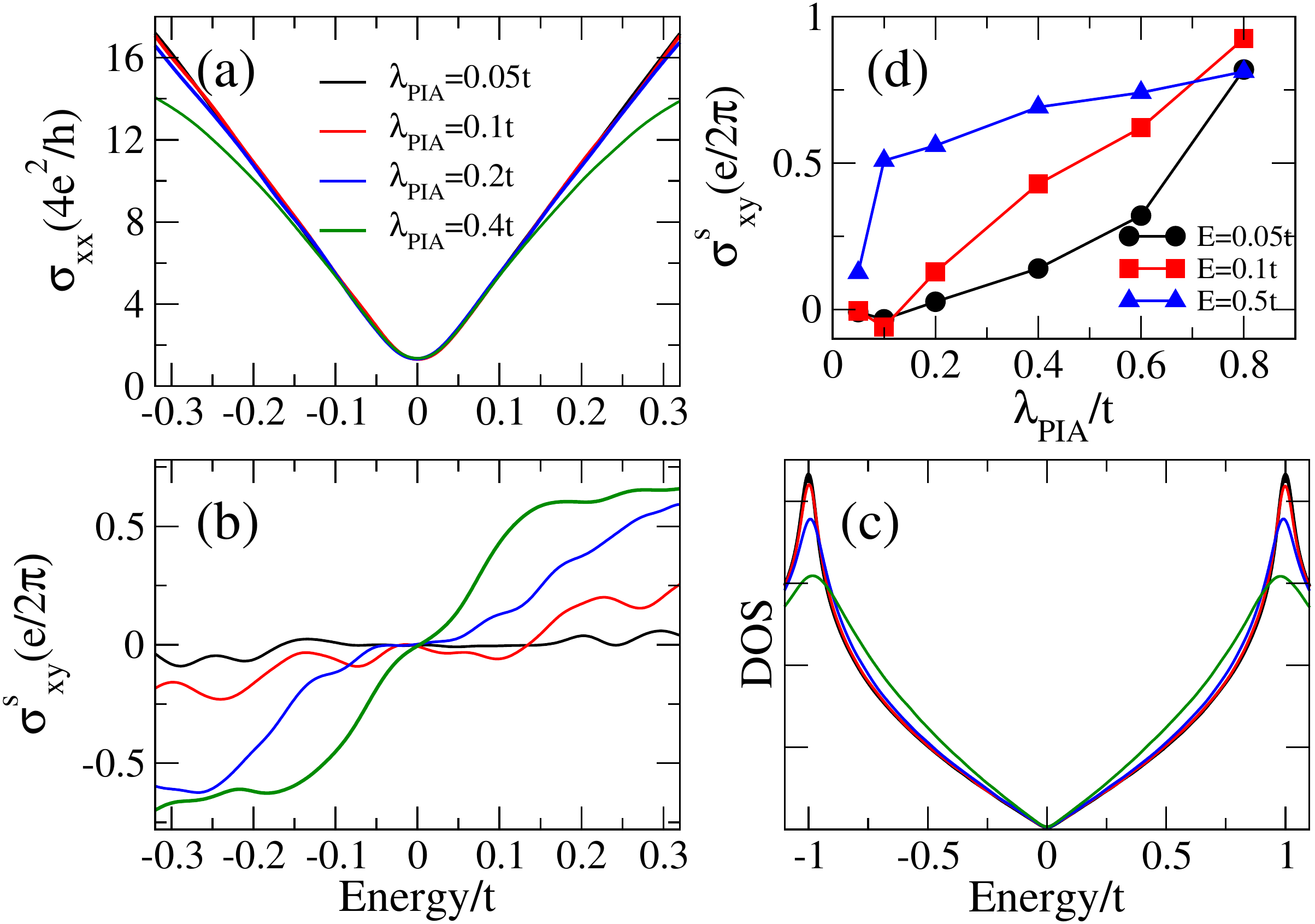}
\caption{ Graphene decorated with a random distribution of PIA-impurities ($\lambda_{I}=\lambda_{R}=0$) sitting at the $T$-site.  $D=2\times 200\times 200$ sites,  concentration of $x=0.2$ and  $M=800$. (a) Longitudinal conductivity  and (b) spin Hall conductivity  as a function of the chemical potential for increasing values of the intensity of the ISO coupling. (c) Density of state  (d) Spin Hall conductivity as a function of the spin-orbit intensity for different values of the Fermi energy: (Black) 0.02t, (Red) 0.05t, (Blue) 0.1t. }\label{Chap5-PIAINT}
\end{figure}

In Fig.\ref{Chap5-RSOCON} we show the results as a function of the concentration, fixing the RSO coupling in $\lambda_{RSO}=0.4t$. Again, the system becomes  progressively more metallic for increasing concentration in the low energy limit (see Fig.\ref{Chap5-RSOCON}.b).  The dependence of the spin Hall conductivity with concentration is shown in Fig.\ref{Chap5-RSOCON}.c, where we see a rapid increase in the spin Hall conductivity In the vicinity of the Dirac point when the concentration is reduced.  Furthermore, Fig.\ref{Chap5-RSOCON}.d indicates a dependency of $1/x$ for the spin Hall  conductivity at a fixed chemical potential close to the neutrality point up to $x=0.05$.  Lower concentrations of impurities, not accessible in our analysis, may affect the spin Hall effect in a different way. Our results indicate that  Rashba tends to delocalize the electrons, inducing a metallic behavior near the Dirac point.  Additionally, they suggest that the spin Hall conductivity generated by Rashba must be relevant in experiments: it is larger for low concentrations and weak Rashba spin-orbit coupling, as it is expected for graphene with adatoms.

\begin{figure}
\centering
\includegraphics[width=\linewidth,clip]{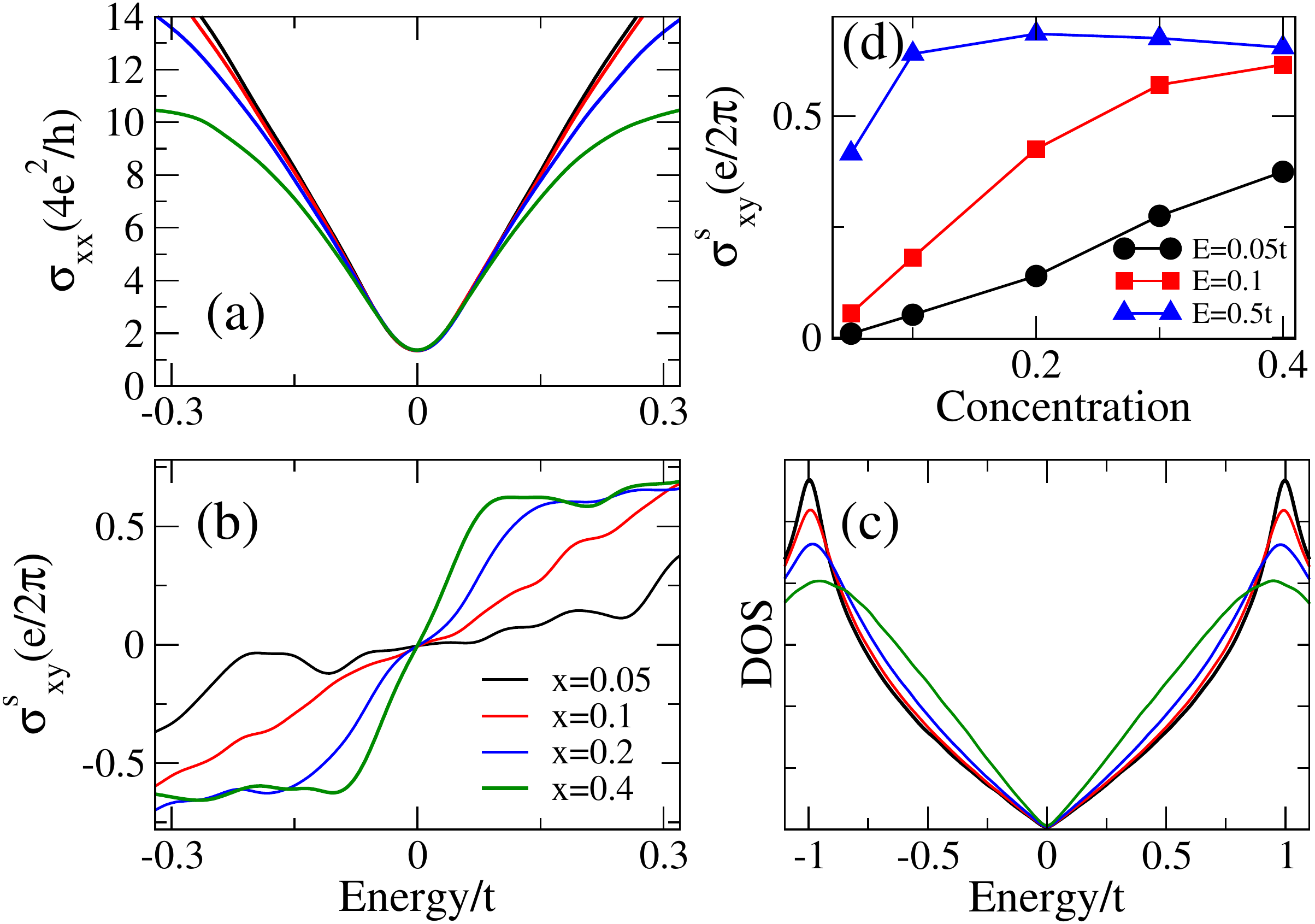}
\caption{ Graphene decorated with a random distribution of PIA-impurities ($\lambda_{I}=\lambda_{PIA}=0$) sitting at the $T$-site.  $D=2\times 200\times 200$ sites,  $\lambda_PIA=0.4t$ and  $M=800$. (a) The longitudinal conductivity  and (b) spin Hall conductivity  as a function of the chemical potential for increasing values of concentration $x$. (c) Density of state  (d) Spin Hall conductivity as a function of the concentration for different values of the chemical potential: (Black) 0.02t, (Red) 0.05t, (Blue) 0.1t.  \label{PIACON}}
\end{figure}

Fig. \ref{Chap5-PIAINT}  presents the results for the pure PIA impurities ($\lambda_{I}=\lambda_{R}=0$) for increasing spin-orbit coupling with $x=0.2$ and $M=800$ . In the density of states, shown in Fig.\ref{Chap5-PIAINT}.c, we can see the emergence of new states in the vicinity of the Fermi energy. These new states are more localized, as can be seen in Fig.\ref{Chap5-PIAINT}.a, where a reduction in the electronic mobility is detected. Fig.\ref{Chap5-PIAINT}.c, displays the spin Hall conductivity  that saturates at high energies with a saturation value that depends directly on the coupling intensity. In Fig.\ref{Chap5-PIAINT}.d we show the behavior of the spin Hall conductivity in the vicinity of the neutrality point. In contrast with the RSO,  for PIA, the spin Hall conductivity increases with the coupling intensity and it is small (tending to zero) in the low coupling regime.

Finally, we consider $\lambda_{PIA}=0.4t$ and display the results as a function of the concentration $x$  (Fig.\ref{PIACON}). In the density of states shown in Fig.\ref{PIACON}.c, we can see again the presence of new localized states in the vicinity of the Fermi energy, which is confirmed by a decrease in the longitudinal conductivity (see Fig.\ref{PIACON}.a). In Fig.\ref{PIACON}.c one can see that the spin Hall conductivity quickly saturates at high energy as a function of the concentration. Near the neutrality point, the behavior is shown in Fig.\ref{PIACON}.d, where the spin Hall conductivity increases with concentration and tends to zero for small concentrations. Our results show that  PIA induces new localized states in the vicinity of the neutrality point. They also indicate that the spin Hall effect  induced by PIA must be negligible under realistic experimental conditions were the SOC is weak and the concentration are small.  
 
 \begin{figure}
\centering
\includegraphics[width=1\linewidth,clip]{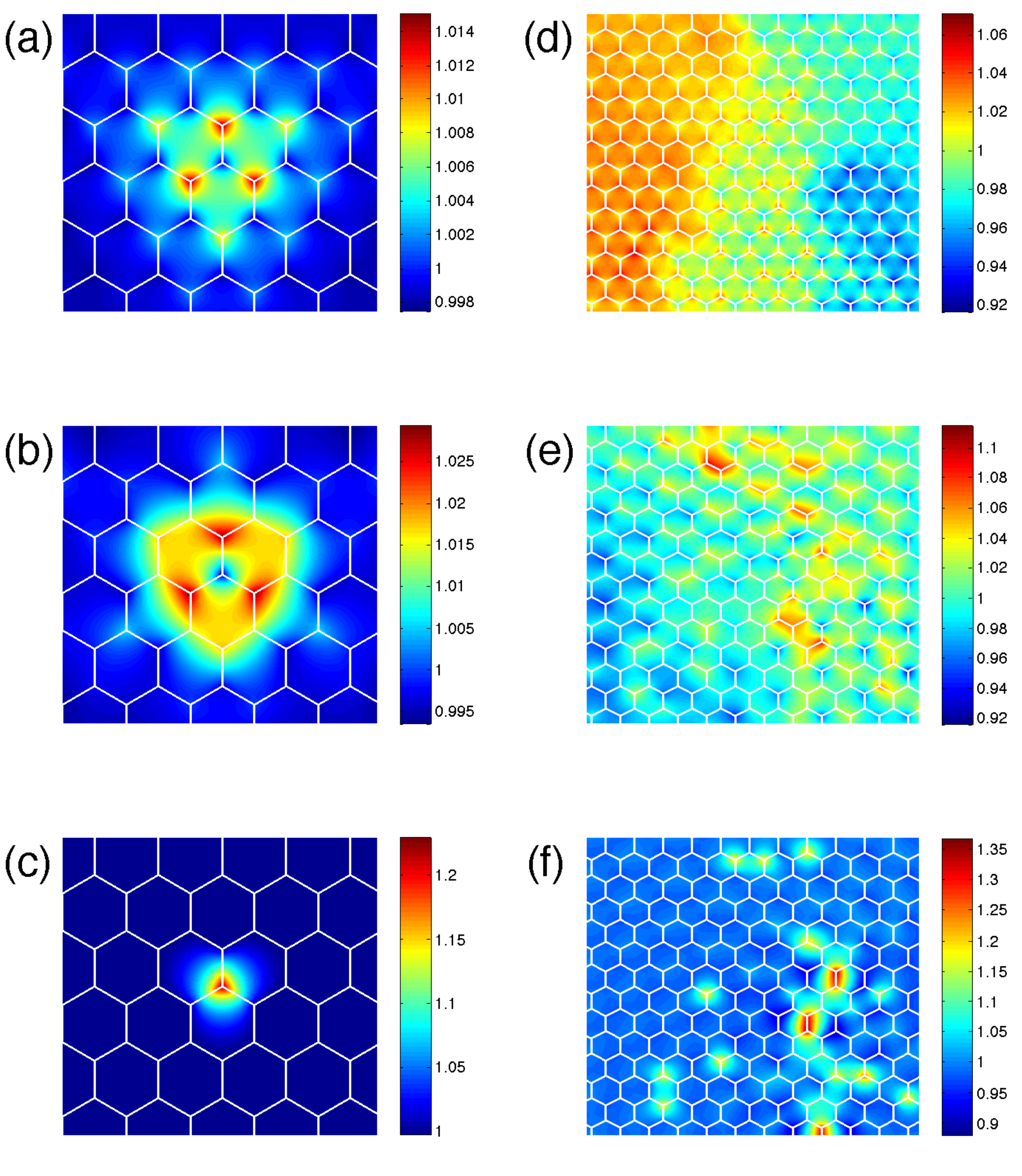}
\caption{Real-space map of the local density of states $\rho_i(E_0)$, near the Fermi level $(E_0=0.05t)$ for : (a) A single ISO impurity $(\lambda_{ISO}=0.4t, \lambda_{RSO}=\lambda_{PIA}=0.0)$, (b) A distribution of ISO impurities with a concentration $x=0.2$ , (c) A single RSO impurity $(\lambda_{R}=0.4t, \lambda_{I}=\lambda_{PIA}=0.0)$, (d) A distribution of RSO impurities with a concentration $x=0.2$,(c) A single PIA impurity $(\lambda_{PIA}=0.4t, \lambda_{ISO}=\lambda_{RSO}=0.0)$, (d) A distribution of PIA impurity with a concentration $x=0.2$}
\label{Real-Space}
\end{figure}

To better understand  how the three spin-orbit couplings affect the electronic properties of graphene, we computed the local density of states LDOS in real-space $\rho_i(\varepsilon)$: 
\begin{equation}
\rho_i(\varepsilon)\equiv\bra{\bm{R}_i}\delta(\varepsilon-H)\ket{\bm{R}_i}=\mp\frac{1}{\pi}\text{Im}\left[G_{i,i}^{\mp}(\varepsilon)\right].\label{LDOSdefinition}
\end{equation}
Following the definition above, $\rho_i(\varepsilon)$ can be considered as the number of available states, for a given energy, at a given lattice site. It can also be expressed as a Chebyshev series and computed numerically~\cite{Garcia2013}.

Figs. \ref{Real-Space}.(a-c) display the effect of an isolated adatom. We can see that the symmetry of each coupling, presented in Fig~\ref{DifferentSpinOrbit} , has a characteristic fingerprint in the local density of states.

Additionally, RSO and ISO couplings have a weak effect on the LDOS with variations of the order of $2\%$  with respect to the mean value. The deviations on the LDOS due to these two terms extends over three lattice constants and there is a reduction of the LDOS at the adsorption site accompanied  by an  increment of the LDOS in nearest and next-next nearest neighbors. PIA has a different behavior: the effect on the LDOS is very localized,  covering a region of only a single lattice constant, with a variation of the LDOS  of the order of $20\%$. 

In Figs. \ref{Real-Space}.(d-f) we show the effects of  a distribution of adatoms for a concentration of $x=0.2$. We can see that for  RSO and ISO impurities, there are small variations of the LDOS at extended regions, which support the previous results where extended states were predicted. For PIA, there are small regions with a concentrated LDOS which is consistent with the picture of states that are more localized. 

\section{Summary}

We presented an extension of the Chebyshev expansion of the Kubo-Bastin formula to calculate the spin Hall conductivity at finite temperature. We applied it to study the effects on spin and charge transport  resulting from adsorbed adatoms that induce spin orbit coupling in graphene . We considered a minimal tight-binding model with three different SOC terms andperformed a systematic analysis of the quantum transport for each of them  in terms of concentration and coupling intensity. 

For the intrinsic spin-orbit coupling, where the appearance of a quantum spin Hall state is expected, we observed a linear dependence of the size of the gap with the concentration of adatoms, for all cases considered here. We also showed the presence of robust spin Hall conductivity outside the topological gap. We found that Rashba induces robust spin Hall conductivity for low concentrations and weak coupling which is an important limit to compare with experimental results. Conversely, PIA  tends to localize electrons and does not contribute to the spin Hall conductivity in the low concentration and weak coupling limit.  In conclusion, both intrinsic and Rashba spin-orbit couplings should be relevant in the regime of parameters that are typically found in experiments. Furthermore, even at finite temperatures and in the presence of disorder, they can give rise to non-quantized but sizable spin Hall conductivities. 

\section*{Acknowledgements}

T.G. R acknowledges the financial support of the Brazilian agency CNPq (Grants No. 477877/2013-3 and 307705/2013-7) and The Royal Society (UK) through a Newton Advanced Fellowship. J. H. G  is grateful to Prof. Alexandre Rocha for hosting him in IFT where this work was completed and also acknowledges the financial support of the Sao Paulo Research Foundation (FAPESP) under grants 2011/11973-4 and 2015/09434-9.

\bibliography{numericalshe}
\end{document}